\documentclass[9pt,twocolumn,twoside]{osajnl}

\usepackage{pgf,tikz}								
\usepackage{pgfplots}
\usepackage{amsmath,amssymb,graphicx}
\usepackage{amsfonts}
\usepackage{amssymb}
\usepackage{Braket}

\newcommand{\vekn}[1]{\mathrm{\textbf{#1}}}
\newcommand{\vekr}{\vekn{r}}
\newcommand{\vekE}{\vekn{E}}

\newcommand{\cl}{\mathrm{c}}
\newcommand{\ci}{\mathrm{i}}

\journal{ol} 

\setboolean{shortarticle}{true} 

\title{Semi-analytical quasi-normal mode theory for the local density of states in coupled photonic crystal cavity-\\waveguide structures}

\author[1,*]{Jakob Rosenkrantz de Lasson}
\author[2]{Philip Trøst Kristensen}
\author[1]{Jesper Mørk}
\author[1]{Niels Gregersen}

\affil[1]{DTU Fotonik, Department of Photonics Engineering, Technical University of Denmark, Ørsteds Plads, Building 343, DK-2800 Kongens Lyngby, Denmark}
\affil[2]{Institut für Physik, Humboldt Universität zu Berlin, Newtonstra{\ss}e 15,  Berlin D-12489, Germany}

\affil[*]{Corresponding author: jakob@jakobrdl.dk}

\dates{Compiled \today}

\ociscodes{(000.3860) Mathematical methods in physics; (050.1755) Computational electromagnetic methods; (050.5298) Photonic crystals; (130.5296)   Photonic crystal waveguides (140.3945); Microcavities; (140.4780) Optical resonators.}

\doi{\url{http://dx.doi.org/10.1364/ol.XX.XXXXXX}}

\begin{abstract}
We present and validate a semi-analytical quasi-normal mode (QNM) theory for the local density of states (LDOS) in coupled photonic crystal (PhC) cavity-waveguide structures. By means of an expansion of the Green's function on one or a few QNMs, a closed-form expression for the LDOS is obtained, and for two types of two-dimensional PhCs, with one and two cavities side-coupled to an extended waveguide, the theory is validated against numerically exact computations. For the single cavity, a slightly asymmetric spectrum is found, which the QNM theory reproduces, and for two cavities a non-trivial spectrum with a peak and a dip is found, which is reproduced only when including both the two relevant QNMs in the theory. In both cases, we find relative errors below $1\%$ in the bandwidth of interest.
\end{abstract}

\setboolean{displaycopyright}{true}

\begin{document}

\maketitle
\thispagestyle{fancy}
\ifthenelse{\boolean{shortarticle}}{\abscontent}{}
Photonic crystals (PhCs), periodic semiconductor systems with sub-micron structuring, are emerging as important building blocks in realizing integrated optical circuits and quantum information networks. Basic PhC elements such as cavities and waveguides are now well-understood and used for photon localization and transport, respectively, and composite systems of one or more PhC cavities and waveguides are being explored for optical switching~\cite{Fan2002,Husko2009,Heuck2013a,Yu2013,Heuck2014}, compact lasers~\cite{Mork2014}, single-photon buffers~\cite{Takesue2013} and optical RAM~\cite{Nozaki2012}. Similarly, at the single emitter-single photon level, coupled PhC cavity-waveguide structures may substantially alter the light-matter interaction~\cite{Yao2010}, which has been demonstrated experimentally~\cite{Englund2007,Schwagmann2012}.

Quasi-normal modes (QNMs) provide a natural and physically appealing basis for the modeling of light in open and leaky resonators~\cite{Ching1998,Kristensen2014}. The QNMs
are solutions to the source-free Maxwell's equations satisfying a radiation condition
and existing at \textit{discrete} and \textit{complex} frequencies, $\tilde{\omega}_\mu = \omega_\mu - \ci \gamma_\mu$. QNMs explicitly account for the leaky nature of the underlying resonator, as quantified by the associated quality factor, $Q_\mu = \omega_\mu/(2\gamma_\mu)$. 

\begin{figure}[t]
\centering
\begin{tikzpicture} [line cap=round,line join=round,x=1.0cm,y=1.0cm, scale=0.30]

\node (img) at (0,12) {\includegraphics[scale=0.175]{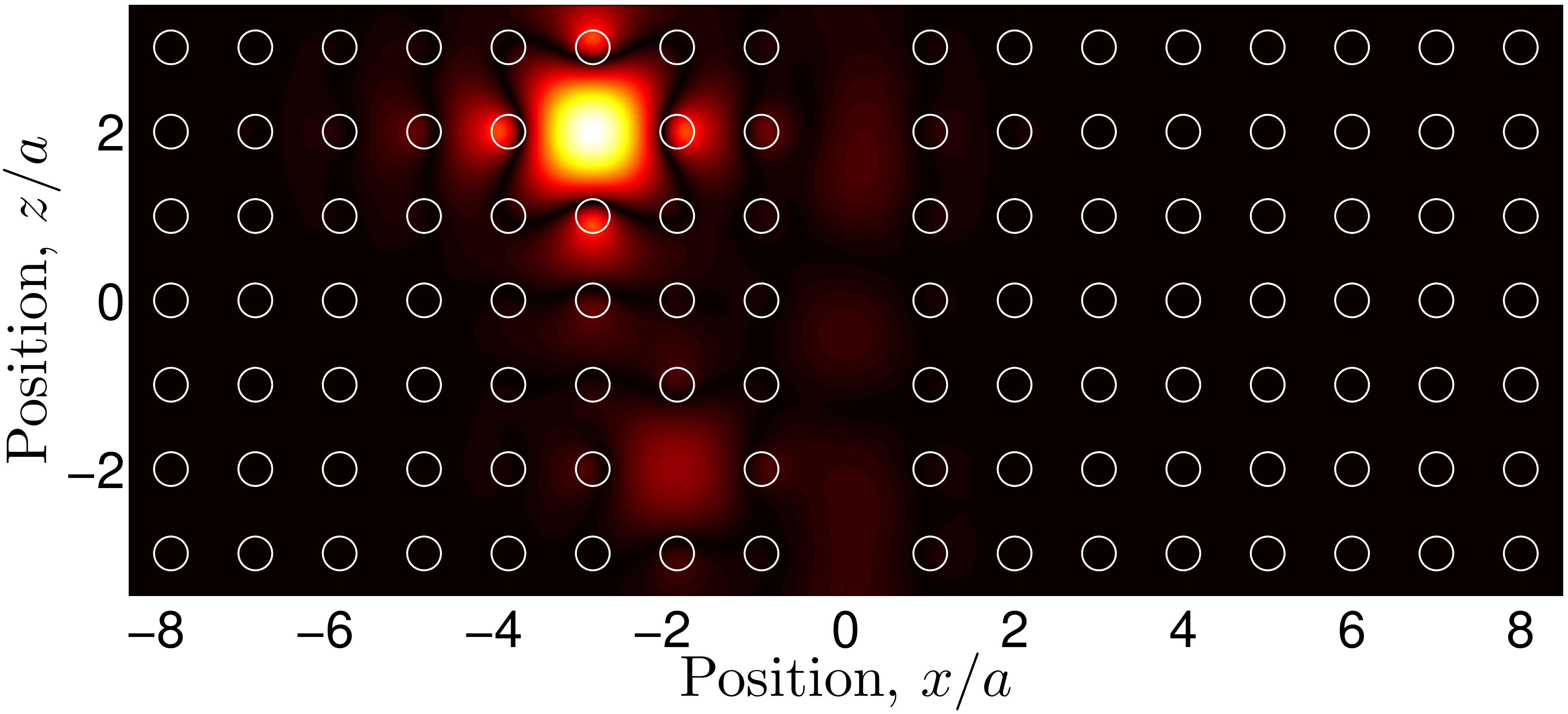}};
\node (img) at (0,0) {\includegraphics[scale=0.175]{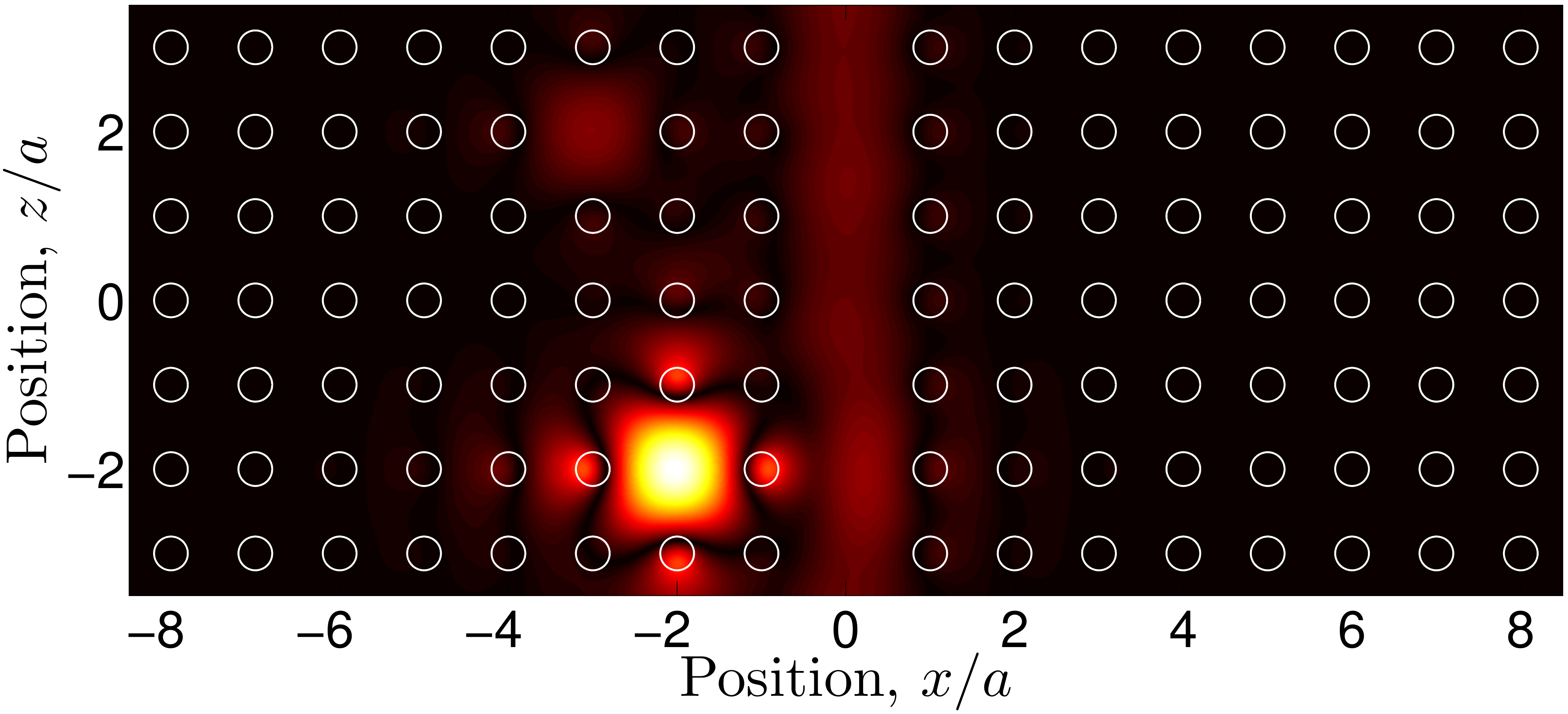}};

\draw [color=green] (14,17) node[circle]{M2};
\draw [color=red] (14,5) node[circle]{M1};

\end{tikzpicture}
\caption{Electric field magnitudes ($|E_y|$) of two QNMs, M1 and M2, for a 2D PhC with two cavities side-coupled to an extended W1 waveguide.}
\label{Fig:Schematic}
\end{figure}
QNM models of optical resonators have been applied to the study of shape perturbations~\cite{Lai_PRA_41_5187_1990,Muljarov_EPL_92_50010_2010} and Green's functions~\cite{Lee1999} in highly symmetric material systems, for which the QNMs can be calculated analytically. Recently, the framework has been applied to more complex dielectric~\cite{Settimi2003,Kristensen2012} and plasmonic~\cite{Sauvan2013,Ge2014a} resonators of practical interest. In these technologically relevant material systems, for which the QNMs must be calculated by numerical means, a description of the optical field in terms of one or a few QNMs has been shown to provide a simple and intuitive, yet surprisingly accurate model. All these successful applications of QNMs share one important characteristic, namely the treatment of resonators embedded in a homogeneous background. Based on the great success of QNM models for such systems, it is natural to ask if the theory can be extended to the technologically relevant case of integrated optical circuits.

In this Letter, we apply the theory of QNMs to set up a semi-analytical model for the (projected) local density of states (LDOS) in systems where optical cavities couple to waveguides that act as the leaky decay channel for the light. To this end, we make use of a regularization of the norm that was recently introduced in order to accommodate the divergent nature of the fields in the waveguides~\cite{Kristensen2014a}. For a given structure, we compute one or a few relevant QNMs and normalize these by a regularization of their divergent far field. Once this is achieved, we can reconstruct the LDOS at any frequency and position in the vicinity of the cavities, which provides intuitive insights into LDOS engineering and is more computationally efficient than full numerical computations. As a proof of principle, we apply the theory to two-dimensional (2D) PhCs where cavities are side-coupled to infinite waveguides. We consider first a single side-coupled cavity (see left inset in Fig.~\ref{Fig:LDOSPhCSideCoupleSingle}) where one QNM provides an accurate description of the LDOS. As a second and more advanced example, the double-cavity structure in Fig.~\ref{Fig:Schematic} is investigated, and we demonstrate that an approximation capturing all features of the LDOS spectrum is obtained only when including both the associated QNMs. The semi-analytical theory is compared to numerically exact computations, and relative errors $< 1\%$ are found, both when one and two QNMs need to be included. Similar configurations with both one~\cite{Fan2002,Husko2009,Heuck2013a} and two~\cite{Heuck2014} side-coupled PhC cavities have also been investigated for optical switching.

In the weak coupling regime, the spontaneous emission rate of a quantum emitter is proportional to the LDOS that, in turn, can be expressed in terms of the dyadic Green's function~\cite{Novotny2012Chap8}
\begin{align} \label{Eqn:LDOS}
\rho^\alpha(\vekr; \omega) = \dfrac{2 \omega}{\pi \cl^2} \mathrm{Im}\left[\vekn{n}_\alpha \cdot \vekn{G}(\vekr, \vekr; \omega) \cdot \vekn{n}_\alpha \right],
\end{align}
where $\vekn{n}_\alpha$ is a unit vector in the direction of the dipole moment of the quantum emitter. Obtaining the LDOS at various positions and frequencies thus amounts to computing $\vekn{G}(\vekr, \vekr; \omega)$, which, unfortunately, can only be done in closed form in a very limited number of simple geometries. In more complex structures, like the PhCs we focus on here, one needs to resort to numerical solvers that are less intuitive and computationally more demanding. As an alternative, we assume that for frequencies close to the cavity resonance frequencies, and at positions in or close to the cavities, $\vekn{G}(\vekr, \vekr'; \omega)$ may be approximated by an expansion on one or a few QNMs. The QNMs are computed and normalized at their discrete frequencies once and for all, and following an approach similar to that of~\cite{Sauvan2013}, for example, one can then expand the Green's function as
\begin{align} \label{Eqn:GQNMExpansion}
\vekn{G}(\vekr, \vekr'; \omega) = \dfrac{\cl^2}{2} \sum_{\mu} \dfrac{\vekE_\mu(\vekr) \otimes \vekE_\mu(\vekr')}{\tilde{\omega}_\mu (\tilde{\omega}_\mu - \omega)},
\end{align}
where $\vekE_\mu(\vekr)$ is the \textit{normalized} electric field of the $\mu$th QNM. By inserting the expression in~\eqref{Eqn:GQNMExpansion} into~\eqref{Eqn:LDOS}, we obtain a semi-analytical QNM representation of the LDOS
\begin{align} \label{Eqn:LDOSQNMExpansion}
\rho^\alpha(\vekr; \omega) = \dfrac{\omega}{\pi} \sum_{\mu} \mathrm{Im} \left[ \vekn{n}_\alpha \cdot \dfrac{\vekE_\mu(\vekr) \otimes \vekE_\mu(\vekr)}{\tilde{\omega}_\mu (\tilde{\omega}_\mu - \omega)} \cdot \vekn{n}_\alpha \right].
\end{align}

In many coupled cavity-waveguide systems of interest
a single or a few QNMs dominate, and retaining only these in the expansion in~\eqref{Eqn:LDOSQNMExpansion} provides a compact and accurate approximation of the LDOS that is more transparent and easier to obtain than a fully numerical computation of the Green's function. Importantly, we do not seek a representation of the Green's function or the LDOS at all positions or frequencies. Therefore, we do not formally rely on a completeness relation for the QNMs, but rather consider the finite sum in~\eqref{Eqn:LDOSQNMExpansion} to be an approximation, which we show below to be extraordinarily good. The work that remains to obtain the LDOS is thus to compute and normalize the QNMs, which we detail in the following.

For resonators surrounded by bulk, the radiation condition that the QNMs must satisfy is the so-called Silver-Müller radiation condition~\cite{Kristensen2014}. When a resonator is coupled to an extended, structured environment, like cavities coupled to an extended waveguide, the Silver-Müller radiation condition is not the correct choice of QNM BC. Instead an outgoing waveguide mode BC must be imposed with only outgoing fields in the waveguide~\cite{deLasson2014b}. This condition can be imposed by use of modal expansion techniques~\cite{Li_JOSA_B_26_2427_2009,deLasson2014b} or by a non-local boundary condition, applicable to standard frequency domain methods~\cite{Kristensen2014a}. Here, we use the modal expansion technique and roundtrip matrix method proposed in~\cite{deLasson2014b} for computing the complex QNM frequencies and associated field distributions. Afterwards, we normalize the QNMs following the procedure in~\cite{Kristensen2014a}, where the spatial integration is split into a finite integration area (volume in 3D) around the cavities and an infinite integration area (volume in 3D) for the extended waveguide. The former contribution is well-behaved and straightforwardly evaluated numerically, while the latter is formally divergent, but can be regularized~\cite{Kristensen2015arXiv} using the theory of divergent series; see all details in~\cite{Kristensen2014a}. This procedure yields the QNM normalization integral, $\Braket{\vekE_\mu | \vekE_\mu}$, defined in~\cite{Kristensen2014a}.

For the specific examples, we consider 2D PhCs, invariant along the $y$ axis, with high-index rods ($\epsilon_{\mathrm{Rods}} = 8.9$, radius $r/a = 0.2$) in a rectangular lattice with lattice constant $a$ and surrounded by air. This structure has a bandgap for the out-of-plane polarization ($\vekE = E_y \hat{\vekn{y}}$), and by leaving out one row of rods a W1 waveguide is created. In the following, we consider two examples of one or more cavities side-coupled to this waveguide; the associated QNMs are leaky due to coupling to the waveguide. In the method from~\cite{deLasson2014b}, we discretize each PhC rod with $N_{\mathrm{R}} = 128$ staircase layers, in which we use $N_{\mathrm{F}} = 101$ Fourier terms in the field expansions. We have checked the accuracy with these parameters and the spatial resolution in computing the QNM normalization integrals, and all numbers are accurate to the quoted number of digits.

As a first example, we consider a single side-coupled cavity positioned a distance $d_{\mathrm{cav}} = 2a$ from the waveguide, see the left inset in Fig.~\ref{Fig:LDOSPhCSideCoupleSingle}. This structure supports a single QNM at $\tilde{\omega}_\mu a /(2\pi \cl) = 0.397 - 0.0014 \ci$, which has an electric field maximum in the center of the side-coupled cavity at $\vekr_{\mathrm{D}}$. The associated complex mode area (volume in 3D) is found to be $a_\mu = \Braket{\vekE_\mu | \vekE_\mu} / (\epsilon(\vekr_{\mathrm{D}}) [\vekE_\mu(\vekr_{\mathrm{D}})\cdot \hat{\vekn{y}}]^2) = (1.441-0.055\ci)a^2$. 
Since this QNM dominates for this structure at $\vekr_{\mathrm{D}}$, it suffices to retain just one term in \eqref{Eqn:LDOSQNMExpansion} of the LDOS QNM expansion
\begin{align} \label{Eqn:LDOSSingleQNM}
\rho^y_{\mathrm{PhC}}(\vekr_{\mathrm{D}}; \omega) = \dfrac{\omega}{\pi } \frac{1}{\epsilon(\vekr_{\mathrm{D}})} \mathrm{Im} \left[ \dfrac{1}{\tilde{\omega}_\mu (\tilde{\omega}_\mu - \omega)} \dfrac{1}{a_\mu} \right],
\end{align}
which can be expressed more explicitly as
\begin{align} \label{Eqn:LDOSSingleQNMExpanded}
\rho^y_{\mathrm{PhC}}(\vekr_{\mathrm{D}}; \omega) &= \dfrac{\omega}{\pi} \frac{1}{\epsilon(\vekr_{\mathrm{D}})} \dfrac{1}{|\tilde{\omega}_\mu|^2} \dfrac{1}{|a_\mu|^2} \dfrac{1}{(\omega - \omega_\mu)^2 + \gamma_\mu^2} \nonumber \\ 
&\hspace{-1.5cm} \times \Big \lbrace \mathrm{Re}\left(a_\mu\right) \left[2\omega_\mu - \omega \right] \gamma_\mu  + \mathrm{Im}\left(a_\mu\right) \left[\omega_\mu \left(\omega - \omega_\mu \right) + \gamma_\mu^2 \right]\Big \rbrace.
\end{align}
This expression is the product of a linear function in $\omega$, a Lorentzian and the term in the curled brackets that depends on the signs and relative magnitudes of $\mathrm{Re}\left(a_\mu\right)$ and $\mathrm{Im}\left(a_\mu\right)$. The expression in \eqref{Eqn:LDOSSingleQNMExpanded} constitutes a semi-analytical single-QNM approximation to the LDOS at $\vekr_{\mathrm{D}}$ for the PhC structure considered here. The associated bulk LDOS is~\cite{Martin1998} $\rho_{\mathrm{Bulk}}^y(\omega) = \omega / (2\pi \cl^2)$, and evaluating the LDOS approximation in \eqref{Eqn:LDOSSingleQNMExpanded} on resonance ($\omega = \omega_\mu$), we find the Purcell factor
\begin{align} \label{Eqn:PurcellSingleQNM}
F_{\mathrm{P}}^y \equiv \dfrac{\rho_{\mathrm{PhC}}^y(\vekr_{\mathrm{D}};\omega_\mu)}{\rho_{\mathrm{Bulk}}^y(\omega_\mu)} = \dfrac{1}{\pi^2} \left(\dfrac{\lambda_0}{n(\vekr_{\mathrm{D}})}\right)^2 \dfrac{Q_\mu}{A_\text{eff}},
\end{align}
where $\omega_\mu / \cl = 2\pi / \lambda_0$, and where we discarded a small term $\gamma_\mu^2 \ll \omega_\mu^2$. The effective mode area
(volume in 3D), $1/A_\mathrm{eff} \equiv \mathrm{Re}(1/a_\mu)$, was defined in~\cite{Kristensen2012}, and the expression in \eqref{Eqn:PurcellSingleQNM} shows that the Purcell formula can be rigorously derived within the framework of QNMs when a single of these dominates the Green's function expansion~\cite{Kristensen2012}. Heuristically, it may be appealing to approximate the single-QNM LDOS enhancement with a Lorentzian curve parametrized with the QNM frequency and the Purcell factor
\begin{align} \label{Eqn:PurcellLorentzian}
\dfrac{\rho_{\mathrm{PhC}}^y(\vekr_{\mathrm{D}};\omega)}{\rho_{\mathrm{Bulk}}^y(\omega)} = F_{\mathrm{P}}^y \dfrac{\gamma_\mu^2}{(\omega - \omega_\mu)^2 + \gamma_\mu^2}.
\end{align}
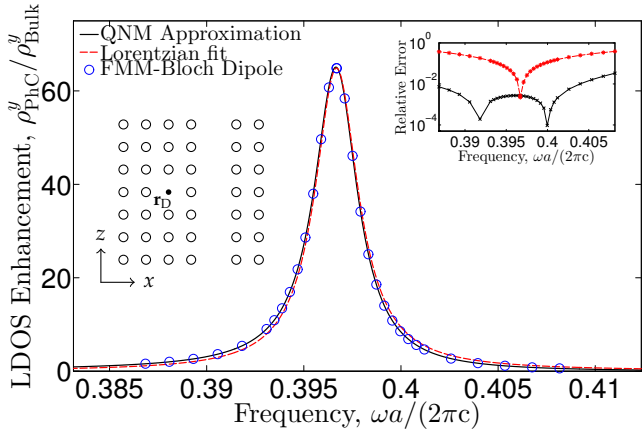
\begin{figure}[htbp]
\centering
\begin{tikzpicture} [line cap=round,line join=round,x=1.0cm,y=1.0cm, scale=0.30]

\node (img) at (0,0) {\includegraphics[scale=0.23]{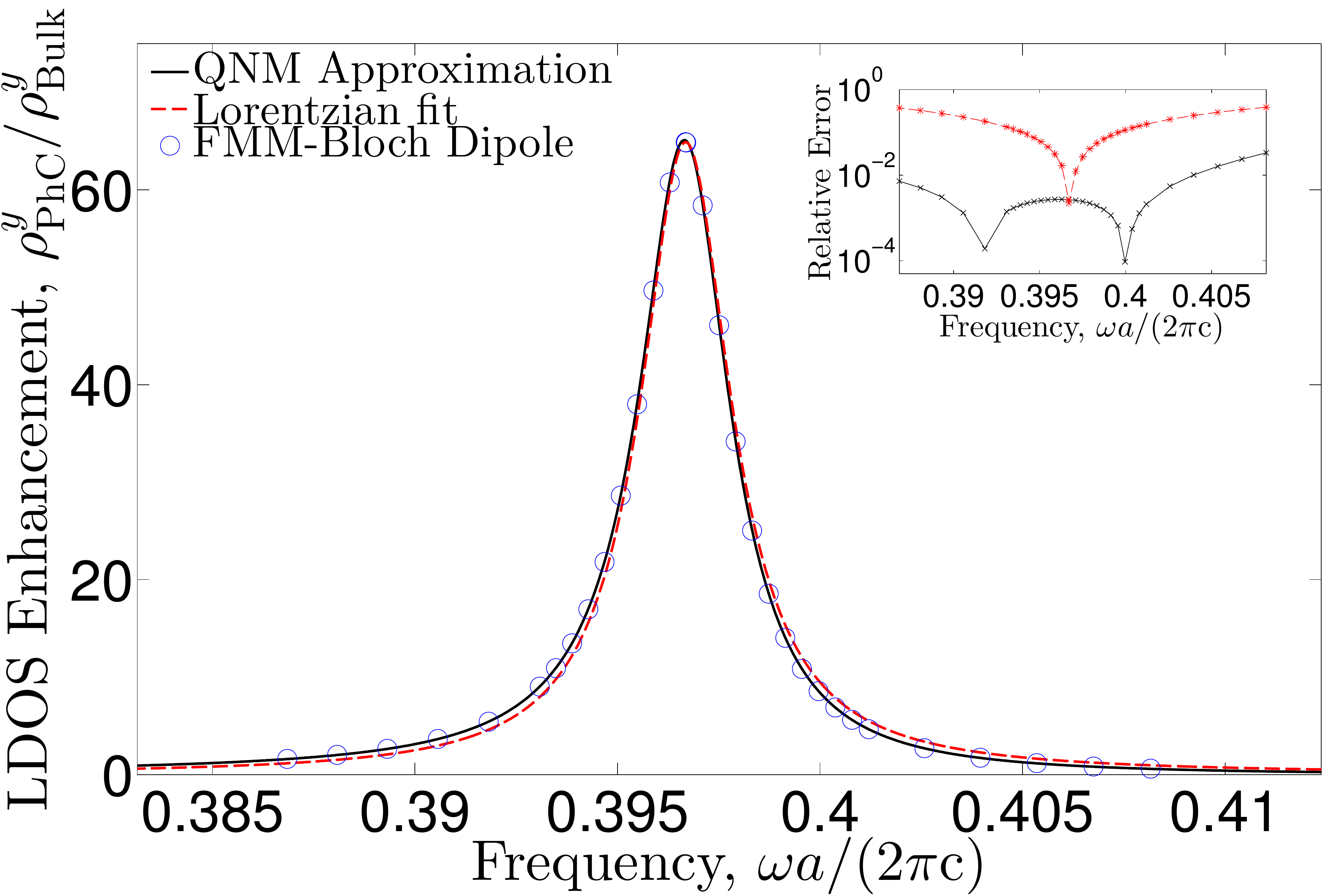}};


Perfect lines of holes
\foreach \i in {1,...,7}
{
	\draw [color=black] (-3+0,-3+\i*1) circle (0.2cm);
	\draw [color=black] (-4+0,-3+\i*1) circle (0.2cm);
	\draw [color=black] (-6+0,-3+\i*1) circle (0.2cm);
	\draw [color=black] (-8+0,-3+\i*1) circle (0.2cm);
	\draw [color=black] (-9+0,-3+\i*1) circle (0.2cm);
}

\foreach \i in {1,2,3,5,6,7}
{
	\draw [color=black] (-7+0,-3+\i*1) circle (0.2cm);
}

\draw [->] (-9.75-0.25,0-3)  -- (-8.25-0.25,0-3) node(xline)[right]{\small $x$};
\draw [->] (-9.75-0.25,0-3)  -- (-9.75-0.25,1.5-3) node(yline)[above]{\small $z$}; 
\fill [color=black] (-7+0,1) circle (3.5pt);
\draw (-7-0.25,2-0.45-1) node[circle]{\footnotesize $\vekr_{\mathrm{D}}$};

\end{tikzpicture}
\caption{Spectrum of LDOS enhancement for a $y$-oriented dipole, $\rho_{\mathrm{PhC}}^y/\rho_{\mathrm{Bulk}}^y$, positioned in the center of a PhC cavity side-coupled at distance $d_{\mathrm{cav}} = 2a$ to a W1 waveguide. The spectrum has been obtained with the single-QNM approximation in \eqref{Eqn:LDOSSingleQNMExpanded} (solid black), with the Lorentzian fit in~\eqref{Eqn:PurcellLorentzian} (dashed red) and with numerically exact 2D FMM-Bloch mode-dipole computations (blue circles). The right inset shows the relative error for the QNM approximation (black) and the Lorentzian fit (red).}
\label{Fig:LDOSPhCSideCoupleSingle}
\end{figure}

In Fig.~\ref{Fig:LDOSPhCSideCoupleSingle}, the solid black curve shows the LDOS approximation in \eqref{Eqn:LDOSSingleQNMExpanded}, while the dashed red curve is the Lorentzian approximation in \eqref{Eqn:PurcellLorentzian}. Numerically exact 2D simulations, obtained using a Fourier modal method (FMM), Bloch mode expansion and S-matrix technique~\cite{Lecamp2007a,Lavrinenko2014Chap6}, are shown as the blue circles. It is apparent that both the single-QNM and the Lorentzian curves approximate the exact spectrum fairly well, but by closer inspection it is also seen that \textit{only} the rigorous single-QNM approxmation (black curve, \eqref{Eqn:LDOSSingleQNMExpanded}) picks up the slight asymmetry of the spectrum, which, by construction, the symmetric Lorentzian does not. Since in this case the real part of $a_\mu$ is much larger than the magnitude of the imaginary part, we can to a good approximation neglect the second term in the curled brackets in \eqref{Eqn:LDOSSingleQNMExpanded}. The slight deviation from the Lorentzian shape of the spectrum thus stems from the first term in the curled brackets, leading to a super (sub) Lorentzian dependence on the red (blue) side of the peak. To be quantitative on the agreement, the right inset in Fig.~\ref{Fig:LDOSPhCSideCoupleSingle} shows the relative errors as function of frequency. Close to the QNM frequency, both approximations provide small errors below 1\%, and while the error from the Lorentzian curve quickly increases away from the resonance, the error from the rigorous expression in \eqref{Eqn:LDOSSingleQNMExpanded} remains below 1\% in most of the considered spectral range. This demonstrates the power of the QNM approach for obtaining accurate LDOS approximations, as also seen in resonators coupled to homogeneous media~\cite{Settimi2003,Kristensen2012,Sauvan2013,Ge2014a}

As a second and more advanced example, we consider the same structure as above, but now add in an additional side-coupled cavity at a distance $d_{\mathrm{cav}} = 3a$ from the waveguide and a distance $d_{\mathrm{W1}} = 4a$ along the waveguide from the initial cavity. This structure supports two QNMs in the spectral range of interest, called M1 and M2, whose electric field magnitudes ($|E_y|$) are shown in Fig.~\ref{Fig:Schematic} where the leakage into the W1 waveguide is clearly visible. The complex QNM frequencies are $\tilde{\omega}_{\mathrm{M1}} a / (2\pi \cl) = 0.397 - 0.0013\ci$ and $\tilde{\omega}_{\mathrm{M2}} a / (2\pi \cl) = 0.395 - 0.00020\ci$, i.e., the two QNMs are offset by approximately 5 nm, while the $Q$ factor for M2 is approximately an order of magnitude larger than that for M1. Since the two QNMs lie relatively close spectrally, and since each QNM has a non-negligible field strength in the adjacent cavity, it is natural to expect that they will both play a role in the QNM-approximated LDOS spectrum. 

\begin{figure}[t!]
\centering 
\begin{tikzpicture} [line cap=round,line join=round,x=1.0cm,y=1.0cm, scale=0.30]

\node (img) at (0,0) {\includegraphics[scale=0.23]{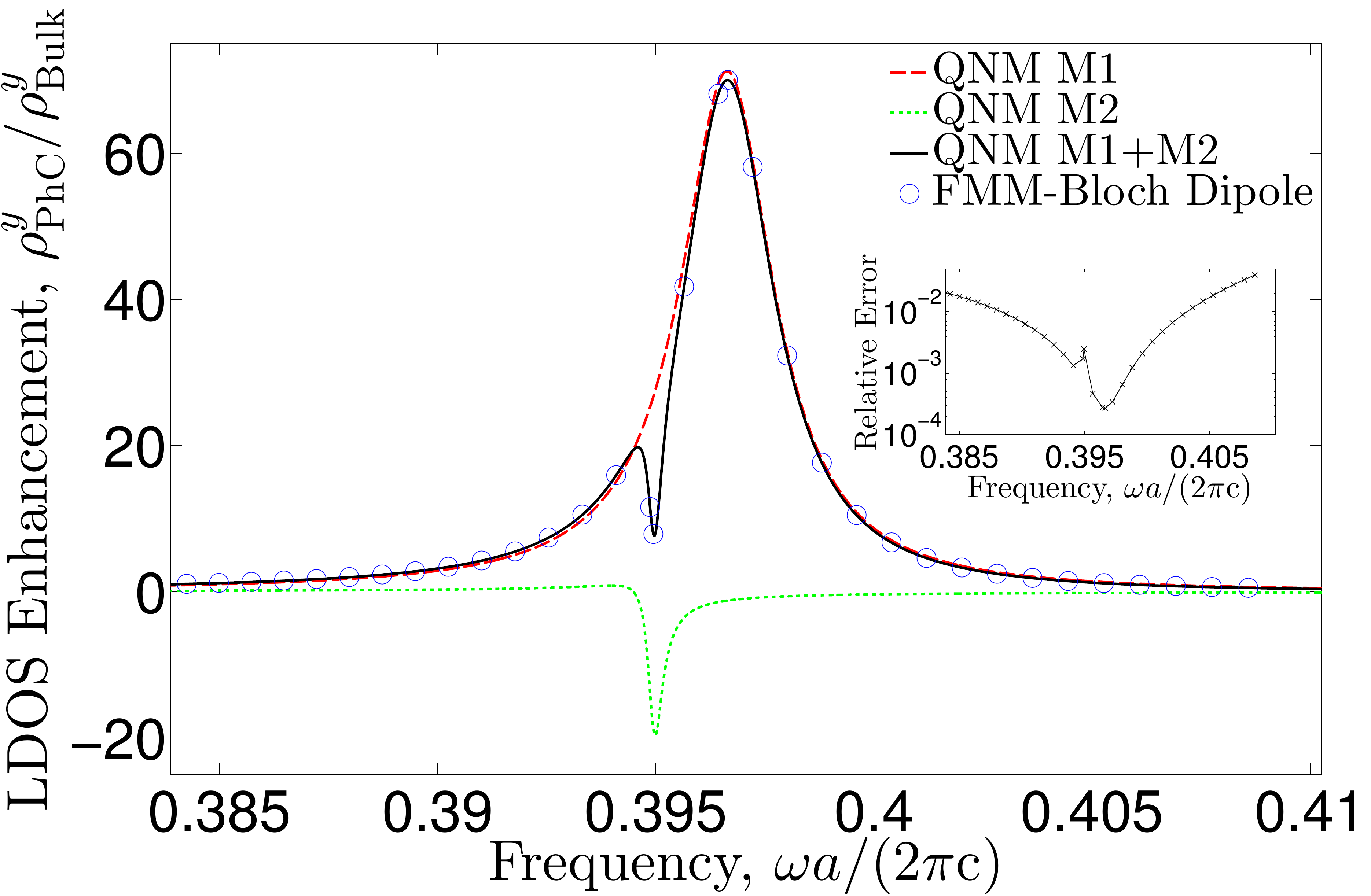}};


\foreach \i in {1,...,7}
{
	\draw [color=black] (-3+1.25,0+\i*1) circle (0.2cm);
	\draw [color=black] (-4+1.25,0+\i*1) circle (0.2cm);
	\draw [color=black] (-6+1.25,0+\i*1) circle (0.2cm);
	\draw [color=black] (-9+1.25,0+\i*1) circle (0.2cm);
	\draw [color=black] (-10+1.25,0+\i*1) circle (0.2cm);
}

\foreach \i in {1,3,4,5,6,7}
{
	\draw [color=black] (-7+1.25,0+\i*1) circle (0.2cm);
}

\foreach \i in {1,2,3,4,5,7}
{
	\draw [color=black] (-8+1.25,0+\i*1) circle (0.2cm);
}

\draw [->] (-9.75,0)  -- (-8.25,0) node(xline)[right]{\small $x$};
\draw [->] (-9.75,0)  -- (-9.75,1.5) node(yline)[above]{\small $z$}; 

\fill [color=black] (-7+1.25,2) circle (3.5pt);
\draw (-7+1,2-0.45) node[circle]{\footnotesize $\vekr_{\mathrm{D}}$};

\end{tikzpicture}
\caption{Spectrum of LDOS enhancement for a $y$-oriented dipole, $\rho_{\mathrm{PhC}}^{y}/\rho_{\mathrm{Bulk}}^{y}$, for the two-cavity configuration in Fig.~\ref{Fig:Schematic} that also shows field profiles of the two QNMs, M1 and M2. The dipole is positioned in the center of the bottom PhC cavity, $\vekr_{\mathrm{D}} = (-2,-2)a$. The spectrum is approximated with a sum over single-QNM terms (\eqref{Eqn:LDOSSingleQNMExpanded}) with the dashed red (dotted green) [solid black] curve obtained with QNM M1 (M2) [M1+M2] included, while numerically exact 2D FMM-Bloch mode-dipole computations are shown as blue circles. The inset shows the relative error for the QNM M1+M2 approximation.}
\label{Fig:LDOSPhCSideCoupleDouble}
\end{figure}
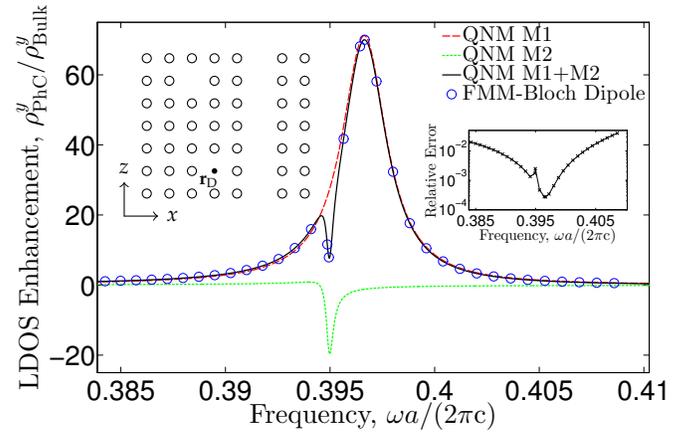

To investigate this structure, we consider a $y$-oriented dipole in the center of the cavity where M1 has its field maximum (see left inset in Fig.~\ref{Fig:LDOSPhCSideCoupleDouble}), and for which we find $a_{\mathrm{M1}} = \Braket{\vekE_{\mathrm{M1}} | \vekE_{\mathrm{M1}}} / (\epsilon(\vekr_{\mathrm{D}}) [\vekE_{\mathrm{M1}}(\vekr_{\mathrm{D}})\cdot \hat{\vekn{y}}]^2) = (1.388-0.026\ci)a^2$ and $a_{\mathrm{M2}} = \Braket{\vekE_{\mathrm{M2}} | \vekE_{\mathrm{M2}}} / (\epsilon(\vekr_{\mathrm{D}}) [\vekE_{\mathrm{M2}}(\vekr_{\mathrm{D}})\cdot \hat{\vekn{y}}]^2) = -(28.7+12.5\ci)a^2$. For both mode areas, the fields are evaluated at the same position, and the emitter is thus spatially offset from the M2 field maximum. We note that for M1 the real part of the complex mode area again dominates and is positive, while for M2 the real and imaginary parts are of the same order of magnitude and both negative. Also, we find that $|\Braket{\vekE_{\mathrm{M1}} | \vekE_{\mathrm{M2}}}| / |\Braket{\vekE_{\mathrm{M1}} | \vekE_{\mathrm{M1}}}| \simeq 10^{-6}$, i.e., QNMs M1 and M2 are orthogonal under the inner product from~\cite{Kristensen2014a}. Using the complex mode areas, we may approximate the LDOS by retaining QNM M1 (dashed red), QNM M2 (dotted green) or QNMs M1+M2 (solid black) in a sum over the single-QNM contribution (\eqref{Eqn:LDOSSingleQNMExpanded}) as shown in Fig.~\ref{Fig:LDOSPhCSideCoupleDouble}. Blue circles again show the numerically exact LDOS enhancement. The exact spectrum features a Lorentzian-like peak close to the M1 QNM frequency and a dip close to the M2 QNM frequency. 

The approximation with only M2 included (dotted green) is negative in a large part of the spectrum, which arises from the negative real and imaginary parts of $a_{\mathrm{M2}}$. Furthermore, while the first term in the curled brackets in \eqref{Eqn:LDOSSingleQNMExpanded} remains negative in the entire spectrum, the second term changes sign at $\omega_{\mathrm{M2}}$, causing the asymmetric lineshape. The approximation with only QNM M1, for which $a_{\mathrm{M1}}$
is dominated by its real part, (dashed red) approximates the peak fairly well, but does not capture the dip close to the M2 frequency. In turn, by including both M1 and M2 (solid black) both features are approximated very well. Close to the M2 frequency the emitter is spectrally (spatially) resonant, but spatially (spectrally) non-resonant with M2 (M1), and we here observe destructive interference between the M1 and M2 terms in the LDOS expansion, which, compared to the single-cavity situation (Fig.~\ref{Fig:LDOSPhCSideCoupleSingle}), lowers the LDOS enhancement. The inset shows that the M1+M2 relative error remains smaller than 1\% in almost all of the considered spectral range, which demonstrates that also when more than one QNM is relevant, the semi-analytical QNM theory proposed here for coupled PhC cavity-waveguide structures is accurate and efficient.

In conclusion, we have demonstrated and validated a semi-analytical quasi-normal mode theory for the local density of states in coupled photonic crystal cavity-waveguide structures. The theory relies on a quasi-normal mode expansion of the Green's function, and once the relevant quasi-normal modes are obtained this expansion gives the Green's function and thus the local density of states at positions and frequencies close to those of the cavities. As a proof of principle, we have demonstrated the theory for two two-dimensional photonic crystal structures where one or two cavities are side-coupled to an extended waveguide. With one cavity, a single quasi-normal mode suffices to approximate the numerically exact LDOS enhancement, with relative errors $< 1\%$, and also picks up a slight asymmetry in the exact spectrum. With two cavities, it is found that two quasi-normal modes are needed to accurately approximate all features of the non-trivial spectrum, and relative errors also here remain $< 1\%$. We foresee that this theory can be useful for analyzing light-matter interactions in more complicated structures, for example including several waveguides~\cite{Davanco2011}, and extension to three-dimensional systems, where full numerical computation of spectra is extremely demanding, should be possible, though also with an increased complexity in the once-and-for-all computation of the relevant quasi-normal modes.

\section*{Acknowledgments}
Support from the Carlsberg Foundation, the Villum Foundation via the VKR Centre of Excellence NATEC and a Sapere Aude Grant LOQIT (DFF-4005-00370) from the Danish Research Council for Technology and Production are gratefully acknowledged. J. R. de Lasson thanks Martijn Wubs for useful discussions.



\begin{thebibliography}{10}
		\newcommand{\enquote}[1]{``#1''}
		
\bibitem{Fan2002}
S.~Fan, \enquote{Sharp asymmetric line shapes in side-coupled waveguide-cavity
	systems,} Appl. Phys. Lett. \textbf{80}, 908--910 (2002).

\bibitem{Husko2009}
C.~Husko, A.~De~Rossi, S.~Combri\'{e}, Q.~V. Tran, F.~Raineri, and C.~W. Wong,
\enquote{Ultrafast all-optical modulation in {GaAs} photonic crystal
	cavities,} Appl. Phys. Lett. \textbf{94}, 021111 (2009).

\bibitem{Heuck2013a}
M.~Heuck, P.~T. Kristensen, Y.~Elesin, and J.~M{\o}rk, \enquote{Improved
	switching using {F}ano resonances in photonic crystal structures,} Opt. Lett.
\textbf{38}, 2466--2468 (2013).

\bibitem{Yu2013}
Y.~Yu, E.~Palushani, M.~Heuck, N.~Kuznetsova, P.~T. Kristensen, S.~Ek,
D.~Vukovic, C.~Peucheret, L.~K. Oxenl{\o}we, S.~Combri\'{e}, A.~de~Rossi,
K.~Yvind, and J.~M{\o}rk, \enquote{Switching characteristics of an {InP}
	photonic crystal nanocavity: Experiment and theory,} Opt. Express
\textbf{21}, 31047--31061 (2013).

\bibitem{Heuck2014}
M.~Heuck, P.~T. Kristensen, and J.~M{\o}rk, \enquote{Dual-resonances approach
	to broadband cavity-assisted optical signal processing beyond the carrier
	relaxation rate,} Opt. Lett. \textbf{39}, 3189--3192 (2014).

\bibitem{Mork2014}
J.~Mork, Y.~Chen, and M.~Heuck, \enquote{Photonic crystal fano laser: Terahertz
	modulation and ultrashort pulse generation,} Phys. Rev. Lett. \textbf{113},
163901 (2014).

\bibitem{Takesue2013}
H.~Takesue, N.~Matsuda, E.~Kuramochi, W.~J. Munro, and M.~Notomi, \enquote{An
	on-chip coupled resonator optical waveguide single-photon buffer,} Nat.
Commun. \textbf{4}, 2725 (2013).

\bibitem{Nozaki2012}
K.~Nozaki, A.~Shinya, S.~Matsuo, Y.~Suzaki, T.~Segawa, T.~Sato, Y.~Kawaguchi,
R.~Takahashi, and M.~Notomi, \enquote{Ultralow-power all-optical ram based on
	nanocavities,} Nat. Photonics \textbf{6}, 248--252 (2012).

\bibitem{Yao2010}
P.~Yao, V.~Manga~Rao, and S.~Hughes, \enquote{On-chip single photon sources
	using planar photonic crystals and single quantum dots,} Laser Photonics Rev.
\textbf{4}, 499--516 (2010).

\bibitem{Englund2007}
D.~Englund, A.~Faraon, B.~Zhang, Y.~Yamamoto, and J.~Vu\v{c}kovi\'{c},
\enquote{Generation and transfer of single photons on a photonic crystal
	chip,} Opt. Express \textbf{15}, 5550--5558 (2007).

\bibitem{Schwagmann2012}
A.~Schwagmann, S.~Kalliakos, D.~J.~P. Ellis, I.~Farrer, J.~P. Griffiths,
G.~A.~C. Jones, D.~A. Ritchie, and A.~J. Shields, \enquote{In-plane
	single-photon emission from a {L3} cavity coupled to a photonic crystal
	waveguide,} Opt. Express \textbf{20}, 28614--28624 (2012).

\bibitem{Ching1998}
E.~S.~C. Ching, P.~T. Leung, A.~Maassen van~den Brink, W.~M. Suen, S.~S. Tong,
and K.~Young, \enquote{Quasinormal-mode expansion for waves in open systems,}
Rev. Mod. Phys. \textbf{70}, 1545--1554 (1998).

\bibitem{Kristensen2014}
P.~T. Kristensen and S.~Hughes, \enquote{Modes and mode volumes of leaky
	optical cavities and plasmonic nanoresonators,} ACS Photonics \textbf{1},
2--10 (2014).

\bibitem{Lai_PRA_41_5187_1990}
H.~M. Lai, P.~T. Leung, K.~Young, P.~W. Barber, and S.~C. Hill,
\enquote{Time-independent perturbation for leaking electromagnetic modes in
	open systems with application to resonances in microdroplets,} Phys. Rev. A
\textbf{41}, 5187--5198 (1990).

\bibitem{Muljarov_EPL_92_50010_2010}
E.~A. Muljarov, W.~Langbein, and R.~Zimmermann, \enquote{{B}rillouin-{W}igner
	perturbation theory in open electromagnetic systems,} Europhys. Lett.
\textbf{92}, 50010 (2010).

\bibitem{Lee1999}
K.~M. Lee, P.~T. Leung, and K.~M. Pang, \enquote{Dyadic formulation of
	morphology-dependent resonances. {I.} {C}ompleteness relation,} J. Opt. Soc.
Am. B \textbf{16}, 1409--1417 (1999).

\bibitem{Settimi2003}
A.~Settimi, S.~Severini, N.~Mattiucci, C.~Sibilia, M.~Centini, G.~D'Aguanno,
M.~Bertolotti, M.~Scalora, M.~Bloemer, and C.~M. Bowden,
\enquote{Quasinormal-mode description of waves in one-dimensional photonic
	crystals,} Phys. Rev. E \textbf{68}, 026614 (2003).

\bibitem{Kristensen2012}
P.~T. Kristensen, C.~V. Vlack, and S.~Hughes, \enquote{Generalized effective
	mode volume for leaky optical cavities,} Opt. Lett. \textbf{37}, 1649--1651
(2012).

\bibitem{Sauvan2013}
C.~Sauvan, J.~P. Hugonin, I.~S. Maksymov, and P.~Lalanne, \enquote{Theory of
	the spontaneous optical emission of nanosize photonic and plasmon
	resonators,} Phys. Rev. Lett. \textbf{110}, 237401 (2013).

\bibitem{Ge2014a}
R.-C. Ge, P.~T. Kristensen, J.~F. Young, and S.~Hughes, \enquote{Quasinormal
	mode approach to modelling light-emission and propagation in nanoplasmonics,}
New J. Phys. \textbf{16}, 113048 (2014).

\bibitem{Kristensen2014a}
P.~T. Kristensen, J.~R. de~Lasson, and N.~Gregersen, \enquote{Calculation,
	normalization, and perturbation of quasinormal modes in coupled
	cavity-waveguide systems,} Opt. Lett. \textbf{39}, 6359--6362 (2014).

\bibitem{Novotny2012Chap8}
L.~Novotny and B.~Hecht, \emph{Principles of Nano-Optics} (Cambridge University
Press, 2012), chap.~8, pp. 224--281.

\bibitem{deLasson2014b}
J.~R. de~Lasson, P.~T. Kristensen, J.~M{\o}rk, and N.~Gregersen,
\enquote{Roundtrip matrix method for calculating the leaky resonant modes of
	open nanophotonic structures,} J. Opt. Soc. Am. A \textbf{31}, 2142--2151
(2014).

\bibitem{Li_JOSA_B_26_2427_2009}
S.~Li and Y.~Y. Lu, \enquote{Efficient method for analyzing leaky cavities in
	two-dimensional photonic crystals,} J. Opt. Soc. Am. B \textbf{26},
2427--2433 (2009).

\bibitem{Kristensen2015arXiv}
P.~T. Kristensen, R.-C. Ge, and S.~Hughes, \enquote{Normalization of
	quasinormal modes in leaky optical cavities and plasmonic resonators,}
(2015). Preprint, arXiv:1501.05938v1 [physics.optics].

\bibitem{Martin1998}
O.~J.~F. Martin and N.~B. Piller, \enquote{Electromagnetic scattering in
	polarizable backgrounds,} Phys. Rev. E \textbf{58}, 3909--3915 (1998).

\bibitem{Lecamp2007a}
G.~Lecamp, J.~P. Hugonin, and P.~Lalanne, \enquote{Theoretical and
	computational concepts for periodic optical waveguides,} Opt. Express
\textbf{15}, 11042--11060 (2007).

\bibitem{Lavrinenko2014Chap6}
A.~V. Lavrinenko, J.~L\ae{}gsgaard, N.~Gregersen, F.~Schmidt, and
T.~S\o{}ndergaard, \emph{Numerical Methods in Photonics} (CRC Press, 2014),
chap.~6, pp. 139--195.

\bibitem{Davanco2011}
M.~Davanço, M.~T. Rakher, W.~Wegscheider, D.~Schuh, A.~Badolato, and
K.~Srinivasan, \enquote{Efficient quantum dot single photon extraction into
	an optical fiber using a nanophotonic directional coupler,} Appl. Phys. Lett.
\textbf{99}, 121101 (2011).

\end{thebibliography}

\pagebreak
\section*{Informational fifth page}
In this section, we provide full versions of citations to assist reviewers and editors.

\end{document}